# PROPAGATION OF OPTICAL SOLITONS IN THE DIELECTRIC MEDIUM OF A LIQUID CRYSTAL

*A.V. Kondakova, T.F. Kamalov*

*Moscow Region State University*

*ul. Very Voloshinoi 24, Mytishchi 141014, Moscow Region, Russian Federation*

*E-mail: kamalov@gmail.com*

*Abstract*

**Aim.** Implement a stochastic representation of the wave function for a pair of entangled soliton functions in a liquid crystal. Show the applicability of a special soliton representation of quantum mechanics for modeling real entangled systems.

**Methodology.** The central place in the study is occupied by the method of mathematical modeling. As part of the calculation of stochastics by the method of abstraction and concretization, a detailed mathematical apparatus is given, adapted to the real physical case. A qualitative analysis of the behavior of the material during the propagation of soliton pulses in it is carried out.

**Results.** The main value of the stochastic theory for a system of entangled solitons lies in the possibility of modeling the entangled states of real systems - photons. In the framework of this work, the optical 1D envelopes of solitons in a nematic liquid crystal are considered in approximation to the conditions of a real physical problem.

**Research implications.** The theoretical and/or practical significance lies in the fundamental possibility of modeling real entangled systems based on the constructed stochastic model of entangled solitons and subsequent creation of special applications on its basis. In particular, there will be a prospect of applying quantum teleportation to the problem of propagation of quantum computing for use among the components of quantum computing networks.

*Acknowledgments:* We thank Professor Belyaev V.V. for good guidance and valuable advice that influenced the writing of this work.

## 1. Introduction

The features of optical soliton propagation in a dielectric are currently actively used in the fields of applied use of quantum phenomena [1-4]. When creating new technologies based on the soliton theory, the issue of developing soliton models available for practical implementation, suitable for subsequent modeling of real entangled systems, is relevant. For example, the simplest model of stochastic modeling of qubits using a soliton scheme and without taking into account the propagation medium [5] opens up the possibility of modeling double entangled configurations for the implementation of quantum algorithms. The method of introducing stochastic q-bits into a soliton scheme using a random phase structure is shown in [6]. However, for the direct application of mathematical models in practice, it is necessary to introduce the characteristics of a specific medium for the propagation of entangled solitons at the initial stage. In our case, the dielectric medium of a nematic liquid crystal was chosen. To analyze the features of soliton propagation in a liquid crystal, equations of soliton motion are compiled based on the nonlinear Schrödinger equation in the approximation of a real physical problem. Given a qualitative understanding of the behavior of solitons in a medium, entangled singlet states of two solitons are constructed in a special stochastic representation. This stochastic representation of the wave function allows one to construct entangled states of solitons that model entangled states of photons.

## 2. Occurrence of an optical soliton in a nonlinear medium

Let us consider a nematic liquid crystal in the absence of absorbing additives as a dielectric medium for the propagation of an optical soliton. For solitons to appear, the corresponding equation of motion must be nonlinear. NLC remains the most widely used in nonlinear optics due to its transparency throughout the entire range from ultraviolet to mid-infrared radiation. At the same time, the crystal has a relatively low electron susceptibility and a high birefringence index. In this case, the required nonlinear term in the equations of soliton motion describing the orientation of the LC director can be provided by external electric or magnetic fields. For static solitons, molecular configurations can be obtained from the Lagrange equation obtained from the free energy density. However, in dynamics, the LC molecules will be in constant motion, and the weakening of the orientation of the molecules cannot be ignored. The orientation inertia term, on the contrary, is usually small and can be neglected. Then the resulting equation of motion will be either a sine-Gordon equation or a double sine-Gordon equation with excessive damping:

$$\frac{\partial^2 u}{\partial t^2} + sinu - \frac{\partial u^2}{\partial^2 x} = 0$$

In nonlinear optics, the main example of an intensity-dependent refractive index is the Kerr response:

$$n(I) = n_0 + n_2 I.$$

In the case of a positive value of $n_2$, the exponent increases with increasing light intensity, and in the case of a finite beam, it leads to a lens-like refractive index distribution that is capable of self-focusing the excitation. As a result, a fundamental soliton is formed, which is a spatially lowest-order mode controlled by a self-induced dielectric waveguide. Since the equations for the electric field of the input polarized light E and the optical director angle for NLC are given as:

$$i\frac{\partial E}{\partial t} + \frac{1}{2}\nabla^2 E - \cos 2\phi\, E = 0;$$

$$\nu\nabla^2\phi + q\cos(2\phi + 2\psi) + 2|E^2|\sin(2\phi) = 0,$$

where $q$ is the relative strength of static electricity with respect to the dynamic electric field, and the angle $\psi$ measures the slope of the static electric field to the direction of propagation, then it is possible to formulate the equations of soliton modulation in NLC [8]. Simplifying the general equations of a liquid crystal to the existing physical situation, when $\psi=0$ at $E=0$, we obtain:

$$i\frac{\partial E}{\partial t} + \frac{1}{2}\nabla^2 E + \sin 2\theta\, E = 0;$$

$$\nu\nabla^2\theta + q\sin(2\theta) + 2|E^2|\cos(2\theta) = 0.$$

Taking into account the boundary conditions in the director deviation, when $\theta$, if $\llbracket r^2 = x^2 + y^2 \to \infty$ we obtain the restoration of its previously established slope. In the current limit of the local response of the material, the obtained director equation can be approximated and solved in the given approximation:

$$-q\sin(2\theta) + 2|E^2|\cos(2\theta) = 0;$$

$$\tan(2\theta) = \frac{2|E^2|}{q}.$$

Substituting the obtained solution into the electric field equation leads to a saturable nonlinear Schrödinger equation of the form:

$$i\frac{\partial E}{\partial t} + \frac{1}{2}\nabla^2 E + \frac{2|E^2|E}{(q^2+4|E^4|)^{1/2}} = 0,$$

where the initial condition of the duration of the transmitted beam z=0 is interpreted as a soliton-like pulse of the form:

$$E(0,r) = A\operatorname{sech}\frac{r}{W}.$$

### 3. Optical solitons in a nematic liquid crystal

Using the solution found, we can proceed to calculate the integrals of motion of an optical soliton, describing its configuration inside the liquid crystal. To this end, we will substitute a suitable trial function into the averaged Lagrangian for the nonlinear Schrödinger equation of higher order in the limit $\nu\to 0$:

$$L = ir(E^*E_z - EE_z^*) - r|E_r|^2 + \frac{2r}{q}|E|^4 - \frac{2r}{q^3}|E|^8,$$

here $|E|^8$ is introduced to stabilize the self-focusing soliton and prevent its collapse by taking into account the nonlinearity of the process, and $E^*$ and $E_z^*$ are complex conjugate quantities.

The trial function for a given Lagrangian will describe the functional form of the soliton in a liquid crystal. Using an additional radially symmetric generalization as in [9], we obtain a function of the form:

$$E = asech\frac{r}{w}\exp(i\sigma) + igexp(i\sigma),$$

where the first term reflects the changing soliton momentum, and the second one – the additional out-of-phase interaction of solitons. It is assumed that the function g does not depend on r. Substituting the derived trial function into the original Lagrangian and integrating the resulting expression over r from 0 to ∞, we obtain its average value:

$$\mathcal{L} = -2(a^2w^2I_2 + \Lambda g^2)\sigma' + 2w^2gI_1a' + 4awgI_1w' - 2aw^2I_1g' - a^2I + \frac{2}{q}I_4a^4w^2 - \frac{2}{q^3}I_8a^8w^2, \text{где } \Lambda = \frac{1}{2}l^2.$$

Taking into account the function g at infinity will lose its physical meaning, since the problem of infinite mass will arise. Because of this, it is allowed to consider $g$ only in a region of length $l$, centered around the position of the pulse. Then, due to the two-dimensionality of the nonlinear Schrödinger equation, the pulse will be represented as a disk, limited by the region $r < l^4$.

The following integrals of motion will be found:

$$\int_0^\infty xsech^2xtanh^2xdx = \frac{1}{3}log2 + \frac{1}{6};$$

$$\int_0^\infty xsechxdx = 2C;$$

$$\int_0^\infty xsech^2xdx = log2;$$

$$\int_0^\infty xsech^4xdx = \frac{2}{3}log2 - \frac{1}{6};$$

$$\int_0^\infty xsech^8xdx = \frac{16}{35}log2 - \frac{19}{105},$$

where *C* is the Catalan constant (C = 0.915965594…). Taking into account the variations of the Lagrangian with respect to the parameters *a, w* and *g*, we obtain variational equations of the form:

$$\frac{d}{dt}(I_1 a w^2) = \Lambda g \frac{d\sigma}{dz};$$

$$I_1 \frac{dg}{dt} = \frac{Ia}{2w^2} - \frac{I_4}{q}a^3 + \frac{3I_8}{q^3}a^7;$$

$$I_2 \frac{d\sigma}{dt} = -\frac{I}{w^2} + \frac{3I_4}{q}a^2 - \frac{7I_8}{q^3}a^6;$$

$$\frac{d}{dt}(I_2 a^2 w^2 + \Lambda g^2) = 0.$$

Уравнения движения, выводимые из двумерного нелинейного уравнения Шредингера, имеют аналогичный вид при использовании одномерного. При этом неподвижная точка этих уравнений движений задается на g = 0, с амплитудой *a* и шириной *w*, связанными соотношением:

The equations of motion derived from the two-dimensional nonlinear Schrödinger equation have a similar form when using the one-dimensional one. In this case, the fixed point of these equations of motion is set at $g = 0$, with amplitude *a* and width *w*, related by the relation:

$$w^2 = \frac{qI}{2a^2}(I_4 - \frac{3I_8}{q^2}a^4)^{-1}.$$

According to the theory, the fixed point of a soliton inside a liquid crystal is the center around which the values of the equations of motion oscillate. As a result, it is possible to study the oscillatory behavior of a soliton.

For a stochastic representation of the wave function of solitons in a liquid crystal, we define an auxiliary complex function. Let us assume that the real field describing the soliton particles has the form:

$$\phi(t,r) = \sum_{k=1}^{N} \phi^{(k)}(t,r);$$

$$sup\phi^{(k)} \cap sup\phi^{(k')} = 0 \; при \; k \neq k'.$$

In this case, the conjugate moments of momentum can be represented as:

$$\pi(t,r) = \frac{\partial \mathcal{L}}{\partial \phi_t} = \sum_{k=1}^{N} \pi^{(k)}(t,r), \quad \phi_t = \frac{\partial \phi}{\partial t};$$

$$\frac{\partial H}{\partial t} = i\frac{\partial}{\partial t}\left(r|E_r|^2 - \frac{2r}{q}|E|^4 + \frac{2r}{q^3}|E|^8\right) + \frac{\partial}{\partial r}\left[\frac{1}{2}r(E_r^* E_{rr} - E_r E_r^*) + \frac{2}{q}r|E|^2(EE_r^* - E^* E_r) - \frac{4}{q^3}r|E|^6(EE_r^* - E^* E_r)\right] = \pi(t,r).$$

Again, substituting the trial function, we obtain the desired momentum equation:

$$\pi(t,r) = \frac{\partial}{\partial t}\left(Ia^2 - \frac{2}{q}I_4 a^4 w^2 + \frac{2}{q^3}I_8 a^8 w^2\right).$$

Using the same mathematical calculations, we obtain the energy conservation equation for a propagating soliton inside a liquid crystal. Integrating from 0 to $r$ and then taking into account the initial values of the amplitude and width of the propagating soliton, we obtain:

$$H = I\hat{a}^2 - \frac{2}{q}I_4 \hat{a}^4 \hat{w}^2 + \frac{2}{q^3}I_8 \hat{a}^8 \hat{w}^2 = 0.$$

Then the mathematical description of a soliton with a small amplitude will look like:

$$\hat{a}^6 = -\frac{q^2 I_4 H}{2II_8}, \quad \hat{w}^2 = \frac{qI}{2I_4 \hat{a}^2}, \quad \Lambda = \frac{1}{2}l^2 = \frac{q^3 I_1^2 I}{24 I_2 I_8 \hat{a}^6}.$$

By introducing the Lagrangian field density and defining an auxiliary function that satisfies the normalization condition [7], we can now define a stochastic version of the wave function, taking into account the total number N of independent tests:

$$\varphi = \frac{1}{\sqrt{2}}\left(\nu A + \frac{i}{\nu}\pi\right);$$

$$\hbar = \int dz |\varphi|^2, \quad \hbar \text{ - Planck's constant.}$$

$$\psi_N(t,z) = (\hbar N)^{-\frac{1}{2}} \sum_{j=1}^{N} \varphi_j(t,z).$$

We will show that $\psi_N$ here plays the role of the probability amplitude by calculating the small-volume integral $\Delta V \gg V_0$, where $V_0 \sim \frac{1}{k}$ denotes the correct soliton size:

$$\rho_N = \frac{1}{\Delta V}\int dz |\psi_N|^2; \quad \Delta V \subset R^1.$$

Then, with probability $P = 1 - a\frac{V_0}{\Delta V}$, where $a \sim 1$, the integral will give the solution:

$$\rho_N = \frac{\Delta N}{N \Delta V},$$

where $\Delta N$ is the number of tests for which the centers of soliton particles may be in the used volume $\Delta V$. Let us consider the procedure of measuring some observable $A$, corresponding, according to E. Noether's theorem, to the generator of the symmetry group $\widehat{M}_A$. For example, the angular momentum is associated with the generator of cosmic translation $\widehat{M}_P = -i\nabla$, the angular momentum is associated with the generator of spatial rotation $\widehat{M}_L = J$, etc. As a result of such transformations, the classical observable $A_j$ for the $j$-th test performed in the experiment with a three-dimensional soliton particle will be represented in the form:

$$A_j = \int d^3x \pi_j i \widehat{M}_A \phi_j = \int d^3x \varphi_j^* V \varphi_j,$$

where $\pi_j$ is the generalized impulse in the $j$-th trial, and $\varphi_j$ is the generalized coordinate (field). In our case, $\varphi_j$ corresponds to $A_j$, and $\pi_j$ corresponds to

$$\pi_j = \left(Ia^2 - \frac{2}{q}I_4 a^4 w^2 + \frac{2}{q^3} I_8 a^8 w^2\right)_j.$$

The corresponding average value is calculated as:

$$E(A) \equiv \frac{1}{N}\sum_{j=1}^{N} A_j = \frac{1}{N}\sum_{j=1}^{N}\int d^3x \varphi_j^* \widehat{M}_A \varphi_j = \int d^3x \psi_N^* \hat{A} \psi_N + O\left(\frac{V_0}{\Delta V}\right),$$

where the Hermitian operator $\hat{A}$ denotes $\hat{A} = \hbar \widehat{M}_A$.

Thus, in accordance with the condition of the volumes under consideration ($\frac{V_0}{\Delta V} \ll 1$), we obtain the standard quantum-mechanical rule for calculating the average values of observables. Similar expressions can be found for one-dimensional particles-solitons. For example, the rotation operator has the standard form:

$$(\widehat{S_k})_{lm} = -i\hbar.$$

When considering singlet states of two solitons – a system of entangled solitons with zero spin and momentum, we obtain:

$$\varphi^{(12)}(t, z_1, z_2) = \frac{1}{\sqrt{2}}[\varphi_L(t, -z_1) \otimes \varphi_R(t, z_2) - \varphi_R(t, -z_1) \otimes \varphi_L(t, z_2)].$$

The normalization condition for the resulting equation:

$$\int dz_1 \int dz_2 \varphi^{(12)} * \varphi^{(12)} = \hbar^2$$

corresponds to the obtained expressions of the stochastic wave function of a system of two entangled solitons inside a liquid crystal:

$$\psi_N(t, z_1, z_2) = (\hbar^2 N)^{-\frac{1}{2}} \sum_{j=1}^{N} \varphi_j^{(12)}.$$

**Conclusion**

The main value of the stochastic theory for a system of entangled solitons is the possibility of modeling entangled states of real systems - photons. In this article, we considered optical one-dimensional envelope solitons in a nematic liquid crystal in approximation to the conditions of a real physical problem. A stochastic representation of the wave function for entangled solitons in a dielectric medium of a liquid crystal with a full description of the mathematical apparatus used is implemented. Based on the data obtained, when establishing the exact characteristics of the propagation medium depending on the goals pursued, it will be possible to obtain calculation models for practical application in a number of still unsolved problems of quantum phenomena. Of particular interest is the possibility of reconstructing the wave function of an entangled soliton system:

$$\psi_N(t, z_1, z_2) = (\hbar^2 N)^{-\frac{1}{2}} \sum_{j=1}^{N} \varphi_j^{(12)}.$$

on any entangled pair of real objects. In particular, the proposed stochastic representation will help to resolve the introduction of the phenomenon of quantum teleportation for use among components of quantum computing networks [10].

The result of the conducted study similarly appears to be obvious in the case of applying the compiled model in the area of actual recording of short-range pulses, where there is difficulty in detecting minimum pulse durations regardless of extraneous noise [11-13].

**REFERENCES**

1. Bouwmeester D. et al. Experimental quantum teleportation //Nature. – 1997. – T. 390. – №. 6660. – C. 575-579.

2. Boschi D. et al. Experimental realization of teleporting an unknown pure quantum state via dual classical and Einstein-Podolsky-Rosen channels //Physical Review Letters. – 1998. – T. 80. – №. 6. – C. 1121.

3. Lee R. K., Lai Y., Malomed B. A. Quantum correlations in boundsoliton pairs and trains in fiber lasers //Physical Review A. – 2004. – T. 70. – №. 6. – C. 063817.

4. Rybakov Y. P., Kamalov T. F. Random Solitons Realization of Quantum Mechanics and Stochastic Qubits //arXiv preprint quant-ph/0401168. – 2004.

5. Rybakov Y. P., Kamalov T. F. Entangled solitons and stochastic q-bits //Physics of Particles and Nuclei Letters. – 2007. – T. 4. – №. 2. – C. 119-121.

6. Rybakov Y. P., Kamalov T. F. Probabilistic simulation of quantum states //Proc. SPIE. – 2009. – Vol. 7023. – C. 702307.

7. Reinbert C. G., Minzoni A. A., Smyth N. F. Spatial soliton evolution in nematic liquid crystals in the nonlinear local regime //JOSA B. – 2006. – T. 23. – №. 2. – C. 294-301.

8. Kath W. L., Smyth N. F. Soliton evolution and radiation loss for the nonlinear Schrödinger equation //Physical Review E. – 1995. – T. 51. – №. 2. – C. 1484.

9. Rybakov Y. P., Kamalov T. F. Entangled optical solitons in nonlinear Kerr dielectric //ICONO 2007: Coherent and Nonlinear Optical Phenomena. –

International Society for Optics and Photonics, 2007. – T. 6729. – C. 67291T.
10. Leuchs G. et al. Scheme for the generation of entangled solitons for quantum communication //Journal of Modern Optics. – 1999. – T. 46. – №. 14. – C. 1927-1939.
11. Riedinger R. et al. Remote quantum entanglement between two micromechanical oscillators //Nature. – 2018. – T. 556. – №. 7702. – C. 473-477.
12. Vasilchikova E. N. et al. Optical rotation dispersion of cholestericnematic mixture //Journal of Physics: Conference Series. – IOP Publishing, 2021. – T. 2056. – №. 1. – C. 012030.
13. Pozhidaev E. P. et al. Electro-optical modulation in planar-oriented ferroelectric liquid crystals with a subwavelength spiral pitch //Liquid crystals and their practical use. – 2017. – Vol. 17. – № 4. – C. 90-96.

**INFORMATION ABOUT THE AUTHORS**

*T.F. Kamalov* – PhD, Assoc. Prof., Theoretical Physics Department, Moscow Region State University, https://orcid.org/0000-0002-4349-4747;

*E-mail:* kamalov@gmail.com.

*A.V. Kondakova* – Bachelor student of Moscow Region State University.